Distribution function of charge carriers in electron-phonon systems with spontaneously broken translational symmetry and "vertical dispersion" in ARPES spectra of cuprates

A. E. Myasnikov[1*], E. N. Myasnikov[2], D. V. Moseykin[1], I. S. Zuev[1]

[1] Department of Physics, Southern Federal University, 5 Zorge str., 344090, Rostov-on-Don, Russia
[2] Department of Mathematics, Computer science and Physics, Pedagogical Institute of Southern Federal University, 5 Zorge str., 344090, Rostov-on-Don, Russia

PACS numbers: 74.25.-q, 74.25.Jb, 71.38.-k, 71.38.Mx

We show that spontaneous breaking of the translational symmetry (SBTS) of strongly interacting electron and phonon fields acknowledged in lightly doped cuprates to interpret broad bands in their ARPES and optical conductivity spectra influences the carrier distribution function as charge carrier momentum is no longer certain. We develop appropriate distribution and use it to calculate the carrier concentration in different states in essentially doped systems with SBTS. This allows studying doping and temperature evolution of ARPES spectra of such systems. We reveal that at increasing both temperature or doping polarons concede dominance to bipolarons that results in a shift of the broad band maximum to higher binding energy as it occurs in ARPES spectra of cuprates. At further doping delocalized carriers appear and coexist with bipolarons. We show that intimate property of systems with SBTS is partition of the momentum space between autolocalized and delocalized carriers caused by Pauli exclusion rule. We demonstrate that in ARPES spectra such partition manifests itself as "vertical dispersion" or "waterfalls" universally observed in all the cuprates at essential doping and in undoped parent compounds. Thus, studying the carriers distribution in systems where electron-phonon interaction breaks the translation symmetry leads simultaneously to both types of waterfalls observed in significantly doped and undoped cuprates. The article also represents the methods developed to calculate the band in ARPES spectrum caused by bipolaron photodissociation and the bipolaron binding energy and radius using coherent states for the phonon field characterization.

1. INTRODUCTION

Strong electron-phonon interaction (EPI) in doped cuprates is widely acknowledged now due to interpretation of broad bands in their ARPES and optical conductivity spectra[1-9]. Indeed, strong EPI generates such a deformation of the phonon vacuum that the system state with the SBTS (polaron) is energetically profitable[10] according to early Landau's idea[11]. The translational symmetry recovery during the polaron photodissociation is accompanied by radiation of multiple phonons, their number is controlled by Poisson distribution[7,8]. This results in broad bands in ARPES and optical conductivity spectra of lightly doped cuprates.

The natural question arising is how does a system with broken by EPI translational symmetry evolve with increasing doping. Experimentally one can trace doping evolution of ARPES spectra of cuprates. At increasing doping they demonstrate shift of the broad band maximum position to higher binding energies.[12,13] Interestingly, the ARPES experiment at low temperature on very underdoped sample with the transition temperature 5K demonstrates this process: the spectral weight (Fig.1b of Ref. 12) is smeared over the whole region of energies that corresponds to both polaronic and bipolaronic bands. Further doping results in appearance of the spectral weight near the Fermi surface (Fig.1a in Ref.12, Refs.13,14). This band is narrow (quasiparticle-like) and is obviously caused by delocalized carriers. It coexists with the high-energy broad band presumably caused by bipolarons. However the part of the free-carriers band around its minimum is absent[13,14] that is referenced as "vertical dispersion" or "waterfalls" phenomenon. The similar phenomenon occurs in undoped parent compounds[15] but there the band with the missed part is broad (polaronic) and does not approach the Fermi level, having the maximum near $-0.4 \div -0.5$ eV.

Thus, ARPES spectra of cuprates demonstrate exceptionally interesting evolution with doping. Our aim is to study theoretically the doping evolution of ARPES spectra of systems with SBTS and compare the result with the evolution of experimental spectra of cuprates. This task demands to develop our views to fermions distribution over states that are not necessarily characterized by a certain momentum. It is reminiscent of conventional superconductivity theory development when understanding the role of EPI by Frohlich was not enough to explain stabilization of pairs, and formulation of Cooper's pairing phenomenon was based on taking into account the carriers distribution over different states. Having in mind similar task for doped cuprates one cannot use conventional Fermi distribution over the states with certain wave vector because of essential uncertainty of the autolocalized fermion momentum. Therefore we should construct appropriate distribution. Let us discuss first what quasiparticles it should take into account.

There are two types of polarons occurring at strong EPI. If the conductivity band is wide progressing localization of the carrier due to strong EPI is limited by increase of its kinetic energy. This results in formation of autolocalized state of a carrier (or two carriers) in created by itself (themselves) polarization potential well of the size much larger than the lattice constant. Such (bi)polaron known as large, or Pekar, one can be explored using continuum approximation (although recently there are attempts to explore it in a node approach with taking into account next- and next-next-nearest neighbors "hopping"[16]). It represents two coupled wave packets formed by electron and phonon fields[10] and coherently moves in the crystal. It is also important for the carrier distribution that the velocity of this double wave packet is limited by the maximum group velocity of phonons.[17,10]

In a system with narrow conductivity band limiting the carrier kinetic energy strong EPI results in formation of small, or Holstein, polaron localized in one elementary cell. (Small polaron forms also if the dominated EPI is short-range, but it is not the case for ionic compounds). The mobility of small polarons and bipolarons is very low in comparison with that observed in doped cuprates,[18-20] so that we will limit ourselves to consideration of large (Pekar) polarons and bipolarons and delocalized carriers.

Thus, two well-established experimentally properties of doped cuprates: broad bands in ARPES and optical conductivity spectra and sufficiently high mobility of charge carriers determine a *minimal model* considered below to study collective properties of systems with SBTS in order to compare them with experimentally observed properties of cuprates. It is doped dielectric with strong long-range (Frohlich) EPI resulting in SBTS. Characteristic for such systems are polarons and bipolarons with the radius of several lattice constants. Below we construct the distribution of charge carriers which includes only the energies of the quasiparticles and their size and does not depend on the model in which they are calculated.

The article is organized as follows. Part 2 presents binding energies, radii and phonon vacuum deformation parameters for Pekar bipolaron calculated to use them in further consideration. In part 3 we discuss the properties of the carriers distribution function in systems with SBTS where the fermion momentum is not necessarily certain and construct such distribution with Hibbs method. In particular, it is demonstrated that SBTS caused by EPI always leads to the momentum space partition between autolocalized and delocalized carriers displaying itself in ARPES spectra as "vertical dispersion". Part 4 describes doping and temperature evolution of the carrier concentration in different states obtained using the distribution developed. Then Part 6 uses this information to predict the corresponding doping and temperature evolution of ARPES spectra of a system with SBTS caused by EPI. It turns out to be in good agreement with experimentally observed doping and temperature evolution of ARPES spectra of cuprates, and the agreement is both qualitative and quantitative without any fitting parameters. Part 6 also demonstrates how the partition of the momentum space between autolocalized and delocalized carriers predicted by the distribution function manifests itself in ARPES spectra of essentially doped and undoped cuprates. Besides, Part 4 discusses the conditions of the bipolaron Bose-condensation. Part 5 describes the new analytical method of calculating the bipolaron contribution to the ARPES spectrum used then in Part 6.

## 2. BINDING ENERGY AND RADIUS OF PEKAR BIPOLARON

Let us apply variational method to find the bipolaron ground state energy and parameters of the vector of the ground state. Frohlich Hamiltonian of interacting two charge carriers and phonon field has the form

$$H = \sum_{i=1}^{2} \left( -\frac{\hbar^2}{2m^*} \Delta_{r_i} - \sum_{k} \frac{e}{k} \sqrt{\frac{2\pi \hbar \omega_k}{V \varepsilon^*}} \left[ b_k e^{ik(r_i+R')} + b_k^+ e^{-ik(r_i+R')} \right] \right)$$
$$+ \sum_k \hbar \omega_k b_k^+ b_k + \frac{e^2}{\varepsilon_\infty} \frac{1}{|r_1-r_2|}, \qquad (1)$$

where $m^*$ is the carrier effective mass, $\mathbf{r_i}$ is the radius-vector of the i-th carrier with respect to the center of the polarization potential well characterized by the radius-vector $\mathbf{R}'$, $b_k^+, b_k$ are the operators of creation and annihilation of a quantum in the k-th harmonic of the phonon field, $V$ is the crystal volume, $\varepsilon_0$ and $\varepsilon_\infty$ are static and high-frequency dielectric permittivity, $\varepsilon^*$ is effective dielectric permittivity characterizing the lattice polarizability, $(\varepsilon^*)^{-1} = \varepsilon_\infty^{-1} - \varepsilon_0^{-1}$.

The most profitable energetically type of Pekar bipolaron is a one-center bipolaron (i.e. two carriers localized in one polarization potential well created by themselves) if the carriers correlation due to their interaction is taken into account.[21,22] Therefore we search a vector of the system ground state in the adiabatic approximation in a form

$$|s\rangle = \Psi(\mathbf{r}_1, \mathbf{r}_2) \prod_{\mathbf{k}} |d_{\mathbf{k}}\rangle, \qquad (2)$$

$$\Psi(\mathbf{r}_1, \mathbf{r}_2) = A e^{-\alpha(r_1^2 + r_2^2)} \left[ 1 + \beta(\mathbf{r}_1 - \mathbf{r}_2)^2 \right], \qquad (3)$$

where $\Psi$ is a normalized wave function of the electrons in the bipolaron with the localization parameter $\alpha$ and the correlation parameter $\beta$, $A = B^{-1/2}(2\alpha/\pi)^{3/2}$ is normalizing coefficient, $B = \left(1 + 3\frac{\beta}{\alpha} + \frac{15}{4}\left(\frac{\beta}{\alpha}\right)^2\right)$, $\prod_k |d_k\rangle$ is a product of the state vectors of the phonon field harmonics which are naturally expressed in the coherent states representation as in the Pekar polaron case.[10]

Indeed, SBTS due to EPI results from the fact that at strong EPI the ground state corresponds to the shifted equilibrium positions in the phonon field harmonics as was noticed already by Holstein.[23] Such deformation of the phonon vacuum can be also called phonon condensate.[10] The most convenient mathematical description of the phonon vacuum deformation is achieved using the coherent states representation where the coherent state of the k-th harmonic of the phonon field in the ground state of polaron or bipolaron is characterized by a parameter $d_k = |d_k|e^{i\varphi_k}$. It represents average value of the creation and annihilation operators of the k-th harmonic of the phonon field in the ground state[10]:

$$<b_k> = d_k, <b_k^+> = d_k^* . \quad (4)$$

Applying (4) one can easily minimize the average energy of the system over the phonon field parameters $|d_k|, \varphi_k$ at fixed electronic wave function (that corresponds to adiabatic approximation).[10] The minimization yields (Appendix)

$$|d_\mathbf{k}| = \frac{2e}{k}\sqrt{\frac{2\pi}{V\varepsilon^*\hbar\omega_\mathbf{k}}}\eta_\mathbf{k} , \varphi_\mathbf{k} = -\mathbf{kR}', \quad (5)$$

where $\eta_\mathbf{k}$ is Fourier-transform of the squared electronic wave function (3). After substitution of $|d_k|, \varphi_k$ (5) into the average energy functional the parameters α and β of the electronic wave function and the bipolaron binding energy are found with the variational method (Appendix).

Fig.1 demonstrates calculated binding energies and radii of Pekar bipolaron $E_{bip}$, $R_{bip}$ and polaron $E_{pol}$, $R_{pol}$ as functions of the medium parameters. $R_{bip}$ is evaluated as the radius of a sphere (let us denote its volume $V_0$) containing approximately 90% of the bipolaron charge: $\int\int_{V_0 V_0}|\Psi(\mathbf{r_1},\mathbf{r_2})|^2 d\tau_1 d\tau_2 = 0.9$, $R_{pol}$ is calculated in the similar way. Fig.1 shows that at the medium parameters corresponding to the strong coupling case $R_{pol}$ and $R_{bip}$ are close and $E_{bip}$ is close to $2E_{pol}$ (similarly to the result obtained at classical description of the polarization field[21]), so that only distribution function can help find the carrier concentration in both states.

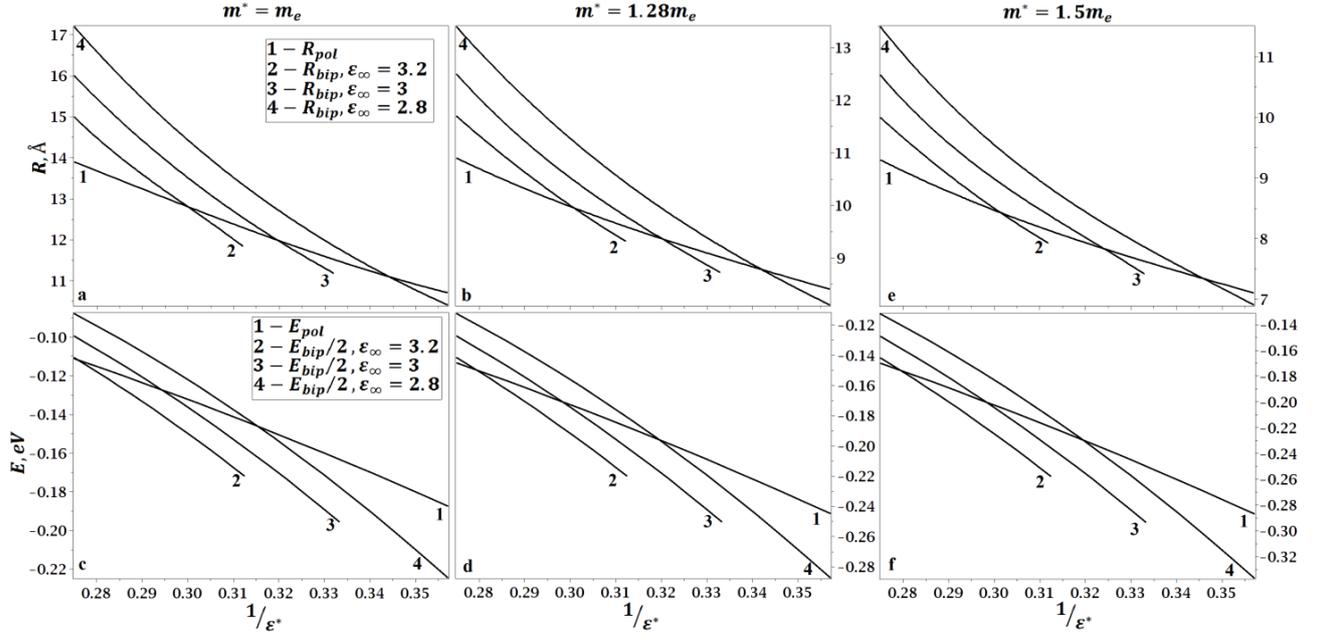

FIG.1.Radii and binding energies of Pekar polaron $R_{pol}$, $E_{pol}$ and bipolaron $R_{bip}$, $E_{bip}$ for $m_e^* = m_e$ (panels a,c), $m_e^* = 1.28 m_e$ (panels b,d) and $m_e^* = 1.5 m_e$ (panels e,f).

## 3. CARRIERS DISTRIBUTION IN SYSTEMS WITH SBTS CAUSED BY EPI

Let us first demonstrate that the necessary distribution should take into account possible coexistence of autolocalized and delocalized carriers.[24] Indeed, let the volume of the bipolaron (or polaron, their volumes are close as Fig.1 shows) is $V_0$. Then the maximum carrier momentum $p_0$ in the (bi)polaron can be estimated as

$$p_0 = 2\pi\hbar/(4\pi V_0/3)^{1/3} . \qquad (6)$$

According to Pauli exclusion rule the carriers added to a system with the carrier concentration higher than $2/V_0$ should have the momentum $p > p_0$. However, the corresponding polaron (bipolaron) velocity $v > p_0/m_{pol}^*$ ($v > p_0/M_{bip}^*$) exceeds significantly the maximum group velocity u of phonons so that the phonon part of the (bi)polaron wave packet cannot follow its electron part in such motion. (Motion of a localized carrier with the velocity v>u generates coherent phonon radiation due to quasi-Cherenkov effect.[10,17,24]) Therefore a crystal of the volume V can contain up to $2V/V_0$ autolocalized carriers, all additional carriers can occupy only delocalized states with $p > p_0$.

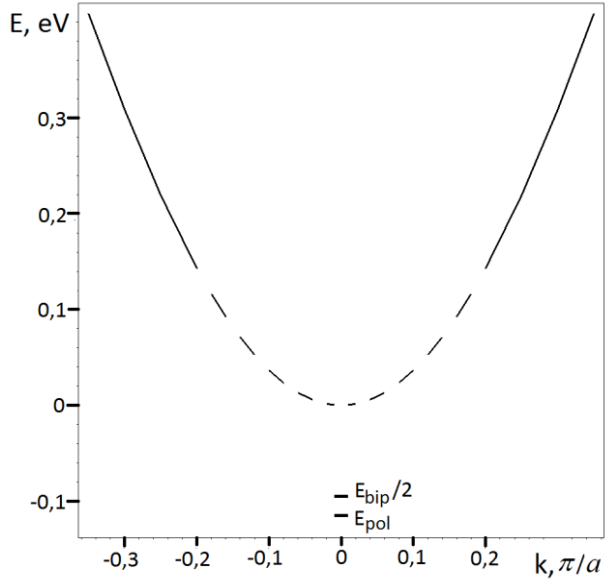

FIG.2. Charge carrier dispersion in a system with SBTS due to EPI. Dashed line demonstrates the part of the carrier band that cannot be filled if all the autolocalized states are occupied. For polaron and bipolaron bands the wave number is proportional to the average carrier momentum.

Thus, strong EPI not only changes the carrier spectrum but, moreover, filling of states of one type depends on the occupation of the others. Fig.2 illustrates this situation for $(\varepsilon^*)^{-1} = 0.28$, $\varepsilon_\infty = 2.85$, $m^* = m_e$, u=$10^4$m/s. Dashed line in Fig.2 demonstrates the part of the carrier band k<$k_0$=$p_0/\hbar$ that cannot be filled if all the autolocalized states are occupied. Since there are no known distribution functions satisfying the conditions described above we deduce it with Hibbs method.

Let us consider possible states of a subsystem occupying the volume $(2\pi\hbar)^3$ in the phase space and localized both in the coordinate space (in the bipolaron volume $V_0$) and in the momentum space (in a sphere of the radius $p_0$ determined by Eq.(6)). Carriers with the momentum $p > p_0$ (called below hot carriers) can be only delocalized so that their concentration $n_{hot}(T,\mu)$, where μ is the chemical potential, is controlled by ordinary Fermi statistics. The (bi)polaron motion does not change essentially the momentum of the carrier contained in it since $u \ll p_0/m_e^*$. Thus, although the carrier momentum and, hence, its energy, in autolocalized states has an uncertainty, the average momentum of the carrier is certain and characterizes the subsystem kinetic energy.

The described subsystem can have occupation numbers N by the carriers 0, 1 or 2 (in the latter case the carrier spins are opposite), so that we should consider it as a grand canonical ensemble. The contribution of the case N=0 to the normalizing condition is obvious. In the case of N=1 the state can be occupied by a polaron or by a delocalized carrier or partially by a polaron and partially by a delocalized carrier, that is taken into account by integration over the filling number from 0 to 1. In the case N=2 the volume $V_0$ can be occupied by a bipolaron or by two interacting polarons or by one polaron and one delocalized carrier or by two delocalized carriers (spins are opposite in all the cases). Besides, the volume $V_0$ can be shared between these four types of states, that is taken into account by integration over the filling number from 0 to 2. The resulting normalizing condition for the distribution of carriers with the average momentums $p < p_0$ (called below cold carriers) has the form

$$e^{\frac{\Omega}{kT}}\left\{1+e^{\frac{\mu}{kT}}\int_0^1 dN_1[N_1 J_1 + (1-N_1)J_2] + e^{\frac{2\mu}{kT}}\int_0^2 dN_1 \int_0^{2-N_1} dN_2 \int_0^{2-N_1-N_2} dN_3\left[\sum_{i=1}^3 N_i I_i + (2-N_1-N_2-N_3)I_4\right]\right\} = 1$$

$$J_1 = \frac{V_0}{(2\pi\hbar)^3} e^{-\frac{E_{pol}}{kT}} \int_0^{m^*u} e^{-\frac{p^2 m^*_{pol}}{2kT m^{*2}}} 4\pi p^2 dp, \quad J_2 = \frac{V_0}{(2\pi\hbar)^3} \int_{m^*u}^{p_0} e^{-\frac{p^2}{2m^* kT}} 4\pi p^2 dp,$$

$$I_1 = \frac{V_0}{(2\pi\hbar)^3} \int_0^{m^*u} e^{-\frac{E_{bip}+p^2 M^*_{bip}/2m^{*2}}{kT}} 4\pi p^2 dp, \quad I_2 = e^{-\frac{E_C}{kT}} J_1^2, \quad I_3 = J_2 J_1, \quad I_4 = J_2^2,$$

where $E_C$ is the energy of Coulomb interaction of two polarons with the same center.

The conventional relation $\bar{n} = -\partial\Omega/\partial\mu$ yields the following expressions for the mean number of carriers in the volume $V_0$ in bipolaron, polaron and cold delocalized states:

$$n_{bip} = \frac{4}{3} A I_1 e^{\frac{2\mu}{kT}}, \quad A = \left[1 + \frac{1}{2} e^{\frac{\mu}{kT}}(J_1 + J_2) + \frac{2}{3} e^{\frac{2\mu}{kT}}(I_1 + I_2 + I_3 + I_4)\right]^{-1} \quad (7)$$

$$n_{pol} = \frac{1}{2} A J_1 e^{\frac{\mu}{kT}} + \frac{4}{3} A(I_2 + I_3/2) e^{\frac{2\mu}{kT}}, \quad n_{coldfree} = \frac{1}{2} A J_2 e^{\frac{\mu}{kT}} + \frac{4}{3} A(I_3/2 + I_4) e^{\frac{2\mu}{kT}},$$

Equation for the carrier chemical potential μ has the form:
$$n_{pol}(\mu,T) + n_{bip}(\mu,T) + n_{coldfree}(\mu,T) + n_{hot}(\mu,T) = n, \quad (8)$$

where n is the total carrier concentration, and for hot delocalized carriers the ordinary Fermi statistics is used. As the solution of Eq.(8) shows the carrier chemical potential in systems with SBTS caused by strong EPI demonstrates unordinary behavior. In particular, due to the limitation of (bi)polaron velocity mentioned above the carrier chemical potential rises with temperature at carrier concentration lower than $2V_0^{-1}$. This results in unusual evolution of the carriers concentration in different states with temperature and total carrier concentration discussed in the next section.

## 4. DOPING AND TEMPERATURE EVOLUTION OF THE CARRIER CONCENTRATION IN DIFFERENT STATES IN ELECTRON-PHONON SYSTEMS WITH SBTS

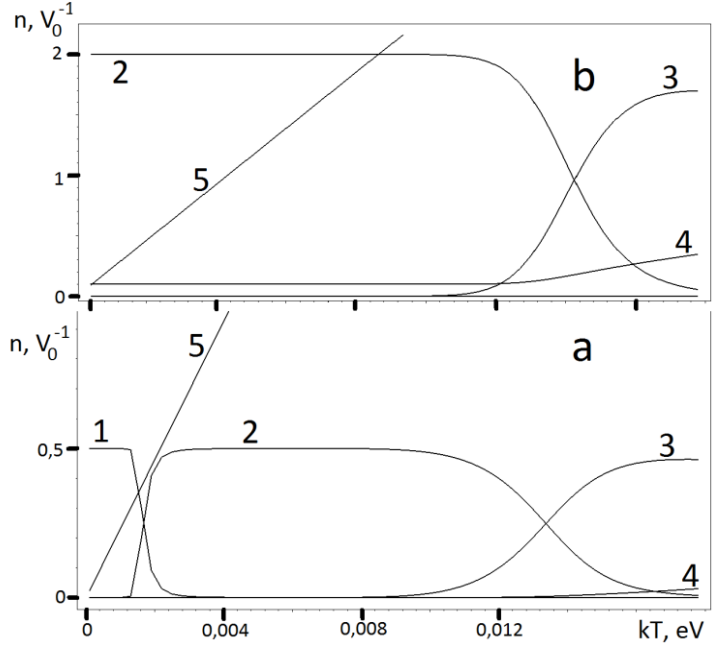

FIG.3. Curves 1,2,3 and 4 display the carrier concentration (in units $V_0^{-1}$) in polaron, bipolaron, cold free and hot free states, accordingly, for the total carrier concentration (a) n=0.5 $V_0^{-1}$ and (b) n=2.1 $V_0^{-1}$. Line 5 shows the maximum carrier concentration in the bipolaron vapor states.

Figs.3a,b demonstrate the carrier concentrations in each state calculated according to (7) as functions of temperature for different total carrier concentration. The medium parameters in Fig.3 are the same as in Fig.2 (except for u=1000 m/s) that yields $E_{pol}$= - 0.115 eV, $E_{bip}$= - 0.196 eV, $E_C$=0.427 eV, $m^*_{pol} = 12 m_e$ (calculated according to Ref. 26), $M^*_{bip} = 24 m_e$ (calculated with a new method developed on the base of Refs.27, 28, 10 and not published yet). Fig.3a shows that in the case $|E_{pol}| > |E_{bip}|/2$ polarons dominate at low temperatures kT< $\Delta E \equiv |E_{pol}| - |E_{bip}|/2$ and carrier concentrations n<1.5 $V_0^{-1}$. At increasing temperature the polarons concede dominance to bipolarons at kT ≈ $\Delta E$ due to the chemical potential growing with temperature. Thus, even in the case $|E_{pol}| > |E_{bip}|/2$ the most part of carriers is in bipolaron states at temperatures T from the interval $\Delta E < kT < kT_0$ where $T_0$ is the bipolaron destruction temperature (the intersection point of curves 2 and 3) depending on their binding energy and maximum group velocity of phonons[29].

Fig3b shows that at carrier concentrations n ≥ 2 $V_0^{-1}$ polarons are absent in the system even if they are more profitable energetically than bipolarons. But at increasing carrier concentration hot delocalized carriers appear in the system at n ≈ 2 $V_0^{-1}$, and their concentration $n_{hot} \approx n - 2V_0^{-1}$ (please, note, that the hot carrier concentration (curve 4) in Fig.3b is 0.1 $V_0^{-1}$ even at zero temperature). Thus, excess carriers added to a system above n=2 $V_0^{-1}$ occupy delocalized carrier states. As simple calculation shows, at the carrier concentration n>2.1$V_0^{-1}$ plasma frequency is higher than phonon frequencies. This leads to screening of EPI that breaks the conditions necessary for the autolocalized carrier states formation. Thus, the developed distribution function is applicable at total carrier concentration n≤2.1$V_0^{-1}$.

Although in this paper our aim is to explain doping evolution of ARPES spectra of cuprates and "vertical dispersion" observed in them on the base of the developed distribution function there is another question that can be considered using the same function. Since for bipolaronic pairing mechanism the superconducting transition temperature is the temperature of Bose-Einstein condensation[18,30] it is interesting to estimate its value. Concentration of bipolarons in Bose-condensate is the difference between the bipolaron concentration and the maximum bipolaron concentration in the vapor. The former (doubled) is shown in Fig.3 with curve 2. The latter we calculate considering bipolarons as Bose-gas with taking into account the described above limitation on the bipolaron momentum:

$$n_{bip.vapor\_max} = \frac{V_0}{(2\pi\hbar)^3} \int_0^{M_{bip}^* u} 4\pi p^2 dp/(e^{p^2/2M_{bip}^* kT} - 1) \quad (9)$$

It (also doubled) is presented in Fig.3 with line 5. The temperature of bipolarons Bose-condensation is abscissa of the intersection point of curve 2 ($n_{bip}$(T)) and line 5 ($n_{bip.vapor\_max}$(T)). As Fig.3 shows it depends on the carrier concentration determining the saturation of curve 2 and on line 5 slope. Optimal doping (the maximum condensation temperature) corresponds to the maximum height of curve 2: $n_{opt}$=2$V_0^{-1}$.

As Fig.3a shows the superconductivity in the case $|E_{pol}| > |E_{bip}|/2$ is absent at low doping (line 5 does not intersect curve 2). However at higher doping when all the carriers are in bipolaron states (as Fig.3 shows $n \approx n_{bip}$ at $n \leq n_{opt}$ and kT> $|E_{pol}| - |E_{bip}|/2$) intersection of curves 2 and 5 appears (in the case presented by Fig.3 at n>0.5$V_0^{-1}$). To simplify (9) one can take advantages of the fact that the maximum kinetic energy of bipolarons is ordinarily much smaller than interesting for us transition temperatures $T_c$ (in the case presented by Fig.3 $M_{bip}^* u^2/2 \approx 10^{-4}$ eV). Then integration in (9) after expanding the exponential yields following relation between the Bose-condesation temperature $T_c$ and carrier concentration n in systems with SBTS due to strong Frohlich EPI:

$$kT_c = \frac{n\pi^2\hbar^3}{M_{bip}^{*2} u}, \quad (10)$$

applicable at $kT_c > M_{bip}^* u^2/2$, $kT_c > |E_{pol}| - |E_{bip}|/2$ and n ≤ $n_{opt}$. In particular, $kT_c^{opt} = \frac{2\pi^2\hbar^3}{V_0 M_{bip}^{*2} u}$. Fig.4 demonstrates the relation between the Bose-condensation temperature and the carrier concentration for the same medium parameters as in Fig.3 in the range of concentrations where the distribution function developed is applicable. Maximum group velocity u of phonons strongly interacting with the charge carrier in Fig.4 is considered independent on the carrier concentration.

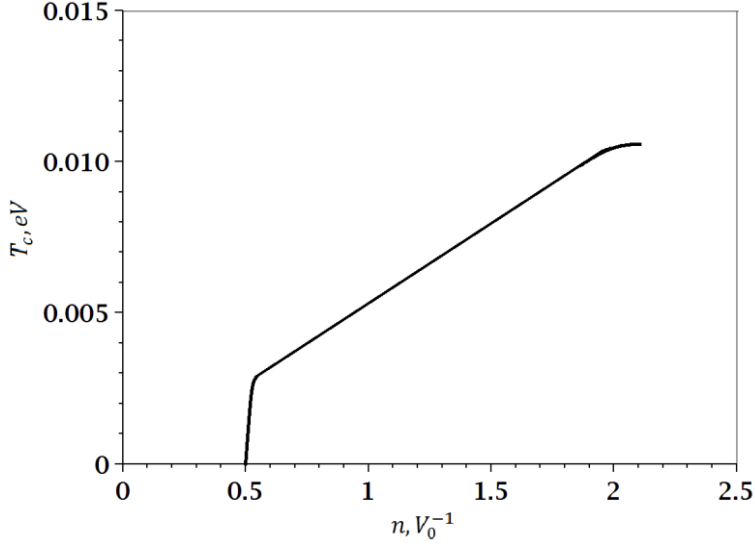

FIG.4. Temperature of bipolarons Bose-condensation as function of the total carriers concentration in the region n$\leq 2.1 V_0^{-1}$ where the developed distribution is applicable.

## 5. BAND IN ARPES SPECTRA CAUSED BY BIPOLARON PHOTODISSOCIATION

Before discussing doping evolution of cuprates ARPES spectra on the base of the information obtained above we should study manifestation of bipolarons in ARPES spectra. Indeed, presence of bipolarons at essential doping is expected on the base of both the distribution function study and the experimentally observed shift of the broad band maximum to higher binding energies with increasing doping.[12,13]

We will use a method [7,8] developed to calculate analytically bands caused by polaron photodissociation in optical conductivity and ARPES spectra. It applies coherent states representation for the phonon field in the polaron and can be easily generalized to describe bipolaron photodissociation. According to Fermi Golden rule the electron phototransition rate into a state with the carrier binding energy from the interval [ε, ε+dε] (it is used instead of the carrier kinetic energy: $\varepsilon = \frac{\hbar^2 k^2}{2m} - \hbar\Omega_{ph}$, where **k** is the photoelectron wave vector inside the crystal and $\hbar\Omega_{ph}$ is the photon energy) and photoelectron wave vector **k** directed in the solid angle dΩ is

$$\frac{dW_{fi}}{dt} = \frac{2\pi}{\hbar} d\varepsilon d\Omega \iint \sum_\nu |\langle \Psi_f|\hat{H}_{int}|\Psi_i\rangle|^2 \delta\left(\varepsilon + \Delta + \nu\hbar\omega + \frac{\hbar^2 k_p^2}{2m_{pol}^*}\right) \frac{V^2 m^* m_{pol}^* k(\varepsilon) k_p}{(2\pi)^6 \hbar^4} d\varepsilon_p d\Omega_p,$$
(11)

where $\Delta = E_{pol} - E_{bip}$, $\varepsilon_p = \frac{\hbar^2 k_p^2}{2m_{pol}^*}$ is the kinetic energy of the arising polaron. We neglect the change of the solid angle dΩ due to the refraction on the material boundary in order to prevent excessively cumbersome formulas. Here we consider one channel of the bipolaron photodissociation – into photoelectron and a polaron. (The second channel of the bipolaron photodisscociation into two electrons yields extremely broad feature that lies much deeper in energy.) The initial state $\Psi_i$ is determined by Eqs.(2,3), the final state is

$$\Psi_f = \frac{1}{V\sqrt{V_0}} \left(\frac{2\pi}{\gamma}\right)^{3/4} e^{-\frac{k_p^2}{4\gamma}} e^{ikr_1} e^{ik_p r_2} |f\rangle$$

where V and $V_0$ are volumes of the crystal and of the polaron, $|f\rangle$ denotes final state of the phonon field combining the phonon vacuum deformation in the polaron and some certain number ν of phonons radiated at the bipolaron photodissociation.

Transverse field of the electromagnetic wave does not affect the longitudinal phonon field in the (bi)polaron. Therefore for the phonon field the probability of the transition from the initial coherent state $|i\rangle$ to the final state $|f\rangle$ where the certain number ν of phonons is present has the form[7,8]

$$P_\nu = |\langle f|i\rangle|^2 = \frac{\bar{\nu}^{\nu-1}}{(\nu-1)!} e^{-\bar{\nu}} \qquad (12)$$

where $\bar{\nu}$ is the average number of phonons radiated during the phonon field relaxation to the final state. For the polaron case $\bar{\nu}^{pol} = \sum_k |d_k^{pol}|^2$, $d_k^{pol}$ is the deformation of vacuum of the k-th phonon field harmonic in the ground state of the polaron[7,8], i.e. in the polaron case the polarization well disappears completely after the photoelectron escape. However, at the bipolaron photodissociation into polaron and photoelectron a part of the phonon vacuum deformation contained in the bipolaron is conserved in the final polaron state. Therefore the average number of radiated phonons $\bar{\nu}$ in (12) is

$\bar{\nu} = \sum_k |d_k^{bip}|^2 - \sum_k |d_k^{pol}|^2$, where $d_k^{bip}$ is the deformation of vacuum of the k-th phonon field harmonic in the bipolaron ground state determined by Eq.(5). $\bar{\nu}$ in (12) can also be calculated as the average energy of the phonon field in the bipolaron and in the polaron, divided by the phonon energy.

The photoelectron kinetic energy is measured experimentally with the resolution determined by the energy window width Δε. Thus, the electron binding energy axis is divided into intervals Δε. The transition probability rate (11) is integrated over $\varepsilon$ inside each interval using δ-function. As the maximum kinetic energy of the polaron is much smaller than the phonon energy $\hbar\omega$ and $\Delta\varepsilon < \hbar\omega$, δ-function argument turns to zero in the only interval Δε for each value of ν. This removes the summation over ν. Integration over the polaron kinetic energy $\varepsilon_p$ due to very small value of the maximum polaron kinetic energy can be carried out approximately, for example, with the simplest rectangles method. At last, integration in (11) over $\Omega_p$ was carried out numerically.

The calculated spectrum is formed by many narrow lines of the width Δε with the centers in the i-th intervals of the carrier binding energy: $\varepsilon_i = -\nu\hbar\omega - \Delta - \overline{\varepsilon_p}$, (where $\overline{\varepsilon_p}$ is the average polaron kinetic energy, ν is integer). The envelope of these narrow lines is determined by Poissonian distribution $P_\nu$ (12). If to take into account phonon dispersion the lines are washed out into a wide band demonstrated by Fig.5. Its maximum lies at the binding energy $\varepsilon_{max} \approx -\Delta - \bar{\nu}\hbar\omega$ (neglecting the average polaron kinetic energy as it is much smaller that the phonon energy). The value $\bar{\nu}$ is always greater than $\bar{\nu}^{pol}$ (determining the maximum of the polaronic band $\varepsilon_{max}^{pol} \approx E_{pol} - \bar{\nu}^{pol}\hbar\omega$) therefore the maximum of the band caused by bipolaron photodissociation corresponds to higher binding energies than the maximum of the band caused by the polaron photoionization. In the present consideration we neglect the reason leading to dispersion of these bands, which will be taken into account elsewhere.

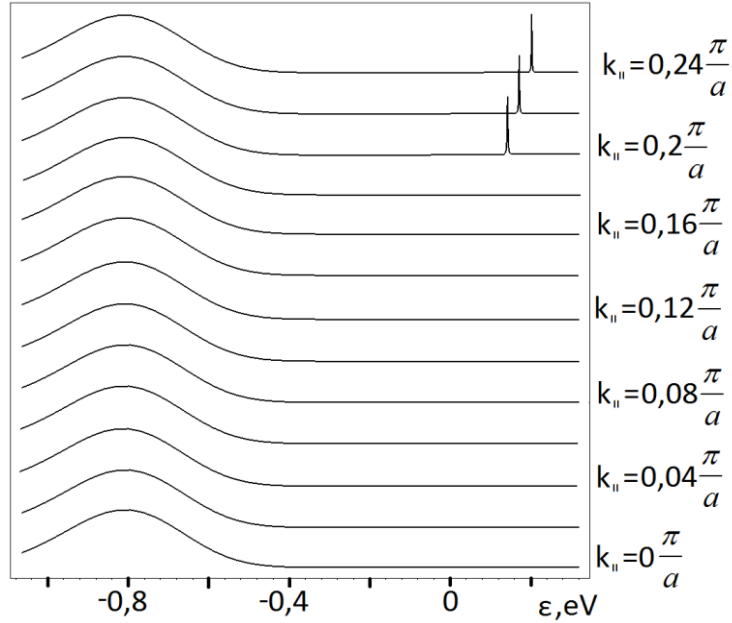

FIG.5. ARPES spectra (energy dispersion curves for different values of the in-plane photoelectron wave vector $k_{||}$ are shifted along the ordinate axis) of the system with SBTS at total carrier concentration n>2$V_0^{-1}$ for the medium parameters $\varepsilon^{*-1}$=0.28, $\varepsilon_\infty$=2.85, $m^*=m_e$. Zero value of ε corresponds to the carriers band extremum.

## 6. DOPING EVOLUTION AND "VERTICAL DISPERSION" IN ARPES SPECTRA OF SYSTEMS WITH SBTS AND IN ARPES SPECTRA OF CUPRATES

Fig3b demonstrates coexistence of polarons and bipolarons at low temperature and coexistence of bipolarons and delocalized carriers in essentially doped systems with strong EPI. As the polarons and bipolarons, autolocalized and delocalized carriers manifest themselves in ARPES spectra in different ways, one can test this prediction studying ARPES spectra of essentially doped cuprates where strong EPI is experimentally proved.[1-9] Besides, ARPES provides a unique opportunity to observe the predicted separation of the momentum space between autolocalized and delocalized carriers demonstrated by Fig.2.

Let us start discussion from the low doping level. At low carrier concentration and temperature kT< $\Delta E \equiv \left|E_{pol}\right|-\left|E_{bip}\right|/2$ the polarons dominate as Fig.3 demonstrates if they are more profitable energetically. In ARPES spectra their photodissociation results in a wide band with the maximum at the binding energy ε≅ -0.4 ÷ -0.5 eV.[8] Such bands are observed in ARPES spectra of underdoped cuprates.[31-33] Undoped parent compounds demonstrate the same bands with the maximum around -0.4 ÷ -0.5 eV.[15] It is natural as the photohole appears in the polaron state. The large width of these bands in ARPES spectra of cuprates is also in agreement with the theoretical prediction of Ref.8. However, the polaron band in undoped compounds demonstrates absence of the low-momentum part, the nature of this phenomenon is discussed below at the end of this section.

At higher temperatures or concentrations polarons concede dominance to bipolarons that is also demonstrated by Fig.3. As was deduced in the previous section the band caused by bipolaron photodissociation has the maximum at higher binding energies than the polaron band. As Figs.5,6 show the bipolaron band maximum can vary considerably depending on the medium parameters, for example, $\varepsilon_{max} \cong -0.8$ eV for $\varepsilon^{*-1}=0.28$, $\varepsilon_\infty=2.85$, $m^*=m_e$, where $R_{bip} \approx 15$ Å (Fig.5), and $\varepsilon_{max} \cong -1.0$ eV for $\varepsilon^{*-1}=0.32$, $\varepsilon_\infty=3.2$, $m^*=1.28m_e$, where $R_{bip} \approx 12$ Å (Fig.6). However, as the polaron band maximum changes in a similar way with the medium parameters[8] and the width of the both bands is large the polaronic and bipolaronic bands will overlap. Therefore at the doping level corresponding to very underdoped sample at temperature lower than kT< $\Delta E \equiv ||E_{pol}|-|E_{bip}||/2$ when polarons and bipolarons can coexist the spectral weight can be extended over the whole region of polaron and bipolaron bands. Such a situation was observed in Ref.12 in the underdoped ($T_c$=5K) BSCCO sample at temperature T=10K as is demonstrated by Fig.1b of Ref.12. At temperatures kT> $\Delta E$ even in the underdoped samples only the bipolaronic band will be observed since at such temperatures polarons are replaced by bipolarons as Fig.3 shows.

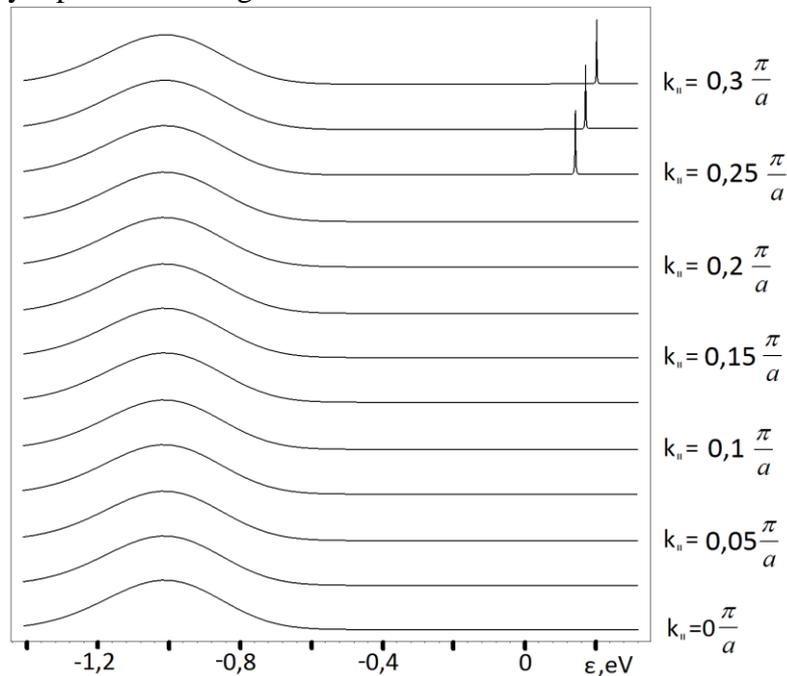

FIG.6. ARPES spectra (energy dispersion curves for different values of the in-plane photoelectron wave vector $k_{||}$ are shifted along the ordinate axis) from the system with SBTS at total carrier concentration n>$2V_0^{-1}$, $\varepsilon^{*-1}=0.32$, $\varepsilon_\infty=3.2$, $m^*=1.28m_e$. Zero value of ε corresponds to the carriers band extremum.

At the doping close to optimal bipolarons dominate, besides hot free carriers emerge as it is illustrated by Fig3b. Hot delocalized carriers, occupying (at essential doping) states k > $k_0$ (solid part of the carrier dispersion in Fig.2), manifest themselves in ARPES spectrum as δ(ε - $E_{free}$(**k**))-function-like peak, where $E_{free}$(**k**) is free carrier dispersion. Since cold delocalized carrier states are not occupied at temperatures T<$T_0$ (where $T_0$ is the bipolaron destruction temperature which can vary from 100 up to 300K) that is illustrated by Fig.3, corresponding to them part k<$k_0$ of the free carrier dispersion (dashed line in Fig.2) is not observed in ARPES spectra. Therefore at k<$k_0$ only the high-energy broad band caused by photodissociation of bipolarons is present in ARPES spectrum whereas at k>$k_0$ narrow band caused by hot delocalized carriers takes place.

Figs.5,6 demonstrate the calculated ARPES spectrum caused by bipolarons and hot free carriers at low temperatures $T<T_0$, when cold free carriers are absent. As Figs.5,6 show the spectrum at n close to $n_{opt}$ consists of two features: deep broad band present at all k values caused by bipolaron photodissociation and narrow peak present only at $k>k_0$ caused by hot free carriers. The latter disperses according to free carriers dispersion from $k=k_0$ until crossing Fermi level. As Figs5,6 demonstrate the obtained display of the *momentum space partition between autolocalized and delocalized carriers* in systems with SBTS caused by EPI both qualitatively and quantitatively coincides with *"vertical dispersion"*, or *"waterfalls"*, observed by several groups on different essentially doped cuprates.[12-14]

Universality of this phenomenon is stressed by observation of "vertical dispersion" in undoped parent compounds,[15] however, the unified interpretation of both these effects was precluded by essential difference between "waterfalls" observed in undoped and lightly doped cuprates. Indeed, although the ARPES spectrum from undoped cuprates also consists of two features[15] and again one of them is the lower-energy band with the absent part $k<k_0$ and the other lies deeper in the energy, the difference from the significant doping case is essential. The low-energy band is so broad that its quasiparticle interpretation is excluded[15], it does not approaches to the Fermi level and has the maximum near -0.45 eV. The higher-energy band is much deeper in energy than in essentially doped cuprates.

Nevertheless, the nature of this phenomenon again can be understood on the base of the momentum space partition between autolocaized and delocalized carriers. The effect was predicted theoretically[34] on the base of a prototype of the present distribution function. It occurs due to absence of EPI screening in a completely filled band. As a result the ARPES spectrum from undoped cuprates consists also of two features.[15] One of them is ordinary polaronic band, which is broad as was discussed above, its maximum disperses near $\varepsilon= -0.45 \div -0.5$ eV. It corresponds to photoemission from delocalized electronic states with the formation of the hole polaron. It is not observed at $k<k_0$ because delocalized initial states are absent at $k<k_0$ as this region of the momentum space is completely occupied by bipolarons. The bipolarons yield the high-energy broad band similar to one shown in Figs.5,6 but with greater (to the valence band width) energy of the maximum since the initial state is electronic bipolaron state lying under the valence band bottom. Upon doping the EPI screening destroys bipolarons at $n \approx 0.05 n_{opt}$ when the plasma frequency becomes comparable with the phonon one as was discussed above.

## 7. CONCLUSION. INFORMATION AVAILABLE FROM ARPES SPECTRA OF SYSTEMS WITH SBTS CAUSED BY EPI

In conclusion, we considered the influence of SBTS caused by strong EPI on the carriers distribution function. We have shown that existing distribution functions are not suitable for such systems and developed the appropriate distribution. It opens an opportunity to describe collective properties of charge carriers in such systems. We apply it to obtain the doping and temperature evolution of the carrier concentration in different states and of ARPES spectra in systems with SBTS due to EPI.

The predicted evolution is in good agreement, both qualitative and quantitative, with the doping and temperature evolution of ARPES spectra of cuprates. In particular, it includes broad band with the maximum near -0.5 eV in undoped compounds and at low doping then coexistence of two bands with the maximum near -0.5eV and near -0.8 ÷ -1.0 eV. At essential and optimal doping there is coexistence of broad bipolaron band with the maximum near - 0.8 ÷ -1.0 eV with narrow peak caused by delocalized hot carriers. Pauli exclusion rule prohibits occupation of delocalized states with the momentums lower than the maximum carrier momentum $p_0$ (6) in the autolocalized state or wave vectors lower than $k_0=p_0/h$. Resulting partition of the momentum space between autolocalized and delocalized carriers inherent in systems where translational symmetry is broken by EPI causes absence of the free carriers spectral weight at $k<k_0$ that looks like "waterfalls" or "vertical dispersion" observed in essentially doped and undoped cuprates.

All described above allows supposing that SBTS caused by EPI widely acknowledged in underoped cuprates (due to interpretation of the broad bands in ARPES and optical conductivity spectra) is preserved also in essentially and optimally doped cuprates and causes characteristic peculiarities in their ARPES spectra. Therefore it is important to discuss what information one can extract from ARPES spectra of systems with SBTS due to EPI. Some of this information is absolutely different from that contained in spectra of metals. First, the maximum and half-width of the band caused by polaron (or bipolaron, depending on the carriers density) photodissociation allows to determine the polaron (bipolaron) binding energy. Second, using the value $k_0$ of the wave vector corresponding to break of the free carrier dispersion in spectra one can calculate the bipolaron volume and radius from (6), where $p_0 = \hbar k_0$. Third, the width of the valence band can be estimated from the energetical distance between the lower and higher energy bands in undoped compound spectrum (Fig.2 in Ref.15).

At last, one can compare the effective masses of the electron in the vicinity of the valence band bottom and of the hole near its top comparing the $k_0$ values in ARPES spectra of undoped (where absence of $k<k_0$ part of low-energy band is caused by electron bipolarons) and essentially doped compound (where the part $k<k_0$ of the momentum space is occupied by hole bipolarons). For example, experiments [12-15] likely show that in cuprates studied there the radius of the electron bipolaron ($k_0 \approx 0.2 \div 0.25$ $\pi/a$ ,[15] that yields according to Eq.(2) $R_{bip} \approx 12 \div 15 \text{Å}$) is larger than that of hole bipolaron ($k_0 \approx 0.25 \div 0.3 \text{Å}^{-1}$ ,[14] $R_{bip} \approx 8 \div 10 \text{Å}$). Such situation is quite possible even in one and the same substance as it is demonstrated by Fig.1 if the electron effective mass is lower than the hole one. Thus, taking into account SBTS caused by EPI in essentially doped and undoped cuprates, not only in the underdoped ones, allows extracting a lot of new information from their ARPES spectra and advancing in understanding their properties.

APPENDIX: CALCULATION OF THE PEKAR BIPOLARON GROUND STATE ENERGY AND GROUND STATE VECTOR PARAMETERS

After a substitution
$$\boldsymbol{R} = \frac{\boldsymbol{r_1} + \boldsymbol{r_2}}{2}, \boldsymbol{r} = \boldsymbol{r_1} - \boldsymbol{r_2}$$
the Hamiltonian (1) and the vector of the system state (2,3) take the form
$$H = -\frac{\hbar^2}{2m^*}\left(2\Delta_r + \frac{1}{2}\Delta_R\right) + \sum_k \hbar\omega_k b_k^+ b_k + \frac{e^2}{\varepsilon_\infty}\frac{1}{r}$$
$$-\sum_k \frac{e}{k}\sqrt{\frac{2\pi\hbar\omega_k}{V\varepsilon^*}}\left[(b_k e^{ikR}e^{ikR'} + b_k^+ e^{-ikR}e^{-ikR'})(e^{ik\frac{r}{2}} + e^{-ik\frac{r}{2}})\right],$$

$$|s\rangle = B^{-1/2}(2\alpha/\pi)^{3/2} e^{-2\alpha R^2} e^{-\alpha \frac{r^2}{2}} (1 + \beta r^2) \prod_k |d_k\rangle$$

Average value of the Hamiltonian in the state $|s\rangle$ is

$$\langle s|H|s\rangle = -\frac{\hbar^2}{2m^*} \int d^3 r \int \Psi^* \left(2\Delta_r + \frac{1}{2}\Delta_R\right) \Psi d^3 R + \sum_k \hbar\omega_k |d_k|^2 + \frac{e^2}{\varepsilon_\infty} \int d^3 r \int \frac{\Psi^2}{r} d^3 R$$

$$- \sum_k \frac{2e|d_k|}{k} \sqrt{\frac{2\pi \hbar \omega_k}{V\varepsilon^*}} \eta_k \left(e^{i(kR'+\varphi_k)} + e^{-i(kR'+\varphi_k)}\right),$$

where

$$\eta_k = \int d^3 R \int \Psi^2 e^{ikR} e^{ik\frac{r}{2}} d^3 r$$

Minimization of the average energy of the system over the phonon field parameters ($|d_k|, \varphi_k$) at fixed electronic wave function yields

$$|d_k| = \frac{2e}{k}\sqrt{\frac{2\pi}{V\varepsilon^* \hbar \omega_k}} \eta_k, \quad \varphi_k = -kR' \quad (A1)$$

After substitution of the phonon field parameters $|d_k|, \varphi_k$ corresponding to the minimum of the functional $\langle s|H|s\rangle$, it takes the form

$$\langle s|H|s\rangle = -\frac{\hbar^2}{2m^*} \int d^3 r \int \Psi \left(2\Delta_r + \frac{1}{2}\Delta_R\right) \Psi d^3 R + \frac{e^2}{\varepsilon_\infty} \int d^3 r \int \frac{\Psi^2}{r} d^3 R$$

$$- \frac{e^2}{\pi^2 \varepsilon^*} \int \frac{\eta_k^{\;2}}{k^2} d^3 k$$

Now we should minimize the bipolaron energy functional $\langle s|H|s\rangle$ over the electronic wave function parameters α and β. Dependence of the energy of bipolaron on the variation parameters has the form

$$E = \delta^2 P_1(\gamma) + \delta(P_2(\gamma) - P_3(\gamma)), \quad (A1)$$

where $\delta = \sqrt{\alpha}, \gamma = \frac{\beta}{\alpha}, P_1(\gamma) = \frac{3\hbar^2}{mB}\left(1 + 2\gamma + \frac{13}{4}\gamma^2\right), P_2(\gamma) = \frac{2e^2}{B\varepsilon_\infty \sqrt{\pi}}(1 + 2\gamma + 2\gamma^2),$

$$P_3(\gamma) = \frac{4e^2}{B^2 \varepsilon^* \sqrt{\pi}}\left(1 + \frac{11}{2}\gamma + \frac{449}{32}\gamma^2 + \frac{2301}{128}\gamma^3 + \frac{43545}{4096}\gamma^4\right).$$

Minimization of Eq.(A1) over the parameter $\delta$ yields $\delta = \frac{P_2 - P_3}{2P_1}$.

Thus, the bipolaron energy as a function of the parameter $\gamma$ has the form $E = -\frac{1}{4}\frac{(P_3(\gamma) - P_2(\gamma))^2}{P_1(\gamma)}$. The first term of Eq.(A1) is the carriers kinetic energy, after substitution of $\delta$ into the term we found $E_{kin} = \frac{1}{4}\frac{(P_3(\gamma) - P_2(\gamma))^2}{P_1(\gamma)}$. Since the result is valid for any value of the parameter $\gamma$, it is also valid for the ground state, so, in the ground state the absolute value of the bipolaron energy is equal to the kinetic energy of the carriers in the bipolaron:

$$|E_{bip}| = E_{kin_0}.$$

Thus, the well-known Pekar theorem 1:2:3:4 proved by him for Pekar polarons is also valid for Pekar bipolarons with the wave function Eq.(3). The subsequent minimization over the parameter $\gamma$ demands solving the equation of the 4-th degree. Therefore it was made numerically, for different values of the medium parameters. The results are depicted in Fig.1.

* e-mail: rochal_s@yahoo.fr